\documentclass{kluwer}
\usepackage{epsfig}
\newdisplay{guess}{Conjecture}
\begin{document}
\begin{article}
\begin{opening}
\title{
Stochastic appearance in stars and high energy sources}
\author{G.S.\surname{Bisnovatyi-Kogan}}
\runningauthor{G.S.Bisnovatyi-Kogan}
\runningtitle{Stochastic appearance}
\institute{Space Research Institute, Russian Academy of Sciences,
Moscow, Russia; gkogan@mx.iki.rssi.ru}
\date{August 15, 2001}

\begin{abstract}

Many phenomena such as stellar variability, stellar explosions,
behavior of different kind of X-ray and gamma-ray sources,
processes in active galactic nuclei and other astrophysical
objects show stochastic features. Brief description of these
phenomena is given in this review.
\end{abstract}

\end{opening}
%\keywords{black holes --- ejection --- active galactic nuclei
%--- stars --- chaos}

\section{Introduction}.
The following phenomena in astrophysics are known to show
stochastic features.

1. Star motion in stellar systems.

2. Active galactic nuclei (AGN)

3. Stellar oscillations

4. X-ray and gamma-ray bursters; quasi-stationary X-ray sources,
like Her X-1.

5. Combustion detonation front in type I supernovae, showing a
fractal structure.

6. Magnetic dynamo in stars.

7. Motion of comets  in Solar system.

Only the first item, star motion in stellar system, is
investigated in details among papers presented in this meeting. In
present paper we shall give a brief review of all the above items.
In particular, for item 1, we consider a
 simple system, consisting of two self-gravitating
intersecting shells with or without the central gravitating body.
The shells consist of stars with the same
parameters of elliptic orbits.

\section{Two shells around SBH.}

Spherically symmetric stellar cluster may be approximated by a
collection of spherical self-gravitating shells consisting of
stars having the same orbit parameters. Dynamical behavior of
a cluster in the shell approximation was first considered by
Yangurazova and Bisnovatyi-Kogan (1984).
Analysis of motion of two such shells consisting of
stars with net radial motion
 with a reflecting inner boundary, done by
Miller and Youngkins (1997), have shown a stochastic behavior
in their motion. More realistic model with two shells of
stars moving along elliptical orbits, with and without the
central gravitating body, was investigated by Barkov et al.
(2001).

The motion of each star in two shells is characterized by the
specific angular momentum $J_1$, $J_2$ which do not change during
intersections, and energies, which are changing during
intersections, and which initial values are

\begin{equation}
 \label{r2}
  E_{1(0)}=\frac{m_1 v_{1(0)}^2}{2}-\frac{Gm_1(M+m_1/2+m_2)}{r} +
  \frac{J_1^2m_1}{2r^2},
\end{equation}

\begin{equation}
 \label{r3}
  E_{2(0)}=\frac{m_2v_{2(0)}^2}{2}-\frac{Gm_2(M+m_2/2)}{r} +
  \frac{J_2^2m_2}{2r^2}.
\end{equation}
Here $v=dr/dt$ is the radial velocity of the shell and
$J^2m/2r^2$ is the total kinetic energy of tangential motions of
all stars of the shell, $M$ is the mass of a central body.
The term $m_1/2$ in (\ref{r2}) is
due to the self-gravity of the shell.
By the index (0) we mark the initial stage before the
first intersection, when the shell "2" is inside the shell "1".
Note, that we use here the values of energies (negative), which
determine the elliptical trajectories of stars in the shells,
including the gravitational energy without normalization.
Let shells
intersect at a some radius $r=a_1$ and at some time
$t=t_1$ after which the shell "1"{} becomes inner and shell "2"{}
-- outer. The "energies" of the shells designated it by the index
(1) than become:

\begin{equation}
 \label{r4}
  E_{1(1)}=\frac{m_1v_{1(1)}^2}{2}-\frac{Gm_1(M+m_1/2)}{r} +
  \frac{J_1^2m_1}{2r^2},
\end{equation}
\begin{equation}
 \label{r5}
  E_{2(1)}=\frac{m_2v_{2(1)}^2}{2}-\frac{Gm_2(M+m_2/2+m_1)}{r} +
  \frac{J_2^2m_2}{2r^2}.
\end{equation}
The matching conditions at the
intersection point $r=a_1,\quad t=t_1$ are written as:

 $$
 E_{1(0)} + E_{2(0)}= E_{1(1)} + E_{2(1)};
 $$
\begin{equation}
  \label{r6}
    v_{1(0)}(t_1)= v_{1(1)}(t_1);\qquad
    v_{2(0)}(t_1)= v_{2(1)}(t_1),
\end{equation}
defining the conservation of the total energy of the system and
continuity of the velocities through the intersection point.
It follows than from (\ref{r2})-(\ref{r6}):

\begin{equation}
  \label{r7}
    E_{1(1)}= E_{1(0)}+\frac{Gm_1m_2}{a_1}; \qquad
   E_{2(1)}= E_{2(0)}-\frac{Gm_1m_2}{a_1}.
\end{equation}
The relations (\ref{r2})-(\ref{r7}) are valid for all subsequent
intersections of the shells.

\begin{figure}
\centerline{\epsfig{figure=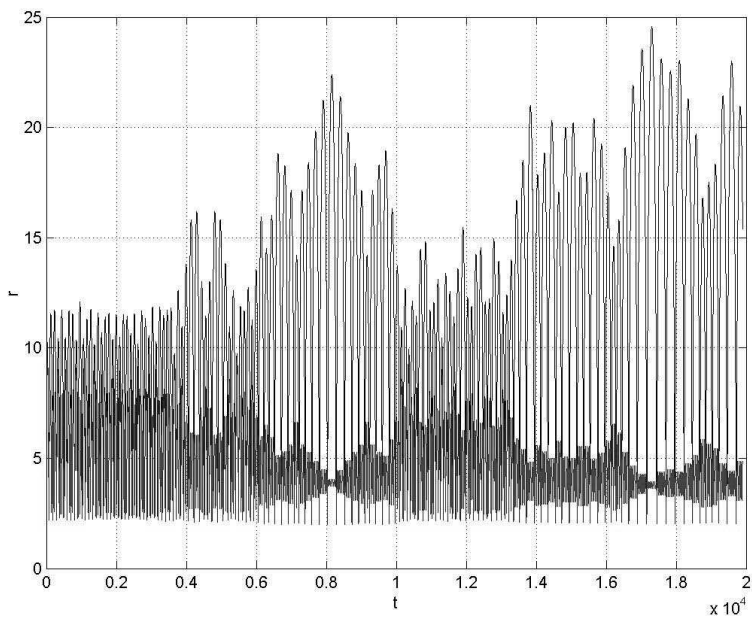,width=12cm}}
\caption{
 The chaotic motion $r_{1,2}(t)$ of shells with equal
masses $m/M=0.08$ on long time interval.}
\label{fig1}
\end{figure}

The shell motion  in the Newtonian gravitational  field is
described by the algebraic relations. The motion of one shell is
completely regular, but at presence of intersections the picture
changes qualitatively. The shell intersections result in chaos in
their motions.
This chaos appears in the fully integrable system, described
algebraically by integrals of motion. The origin of this chaos
is different from the chaotic behavior of non-integrable orbits
in non-axisymmetric gravitational potential (Merritt, 2001).
Character of chaos in the shell motion
depends, mainly, on the mass ratio
of a shell and a  central body. For small mass ratios the motion
of the shells occurs basically in the field of the central body,
and after the intersection there is a little change in a
trajectory of each shell.
However it is possible to observe
randomness of their behavior
in fig. \ref{fig1}.
%Calculations and analysis made by
from Barkov et al. (2001).

\begin{figure}
\centerline{\epsfig{figure=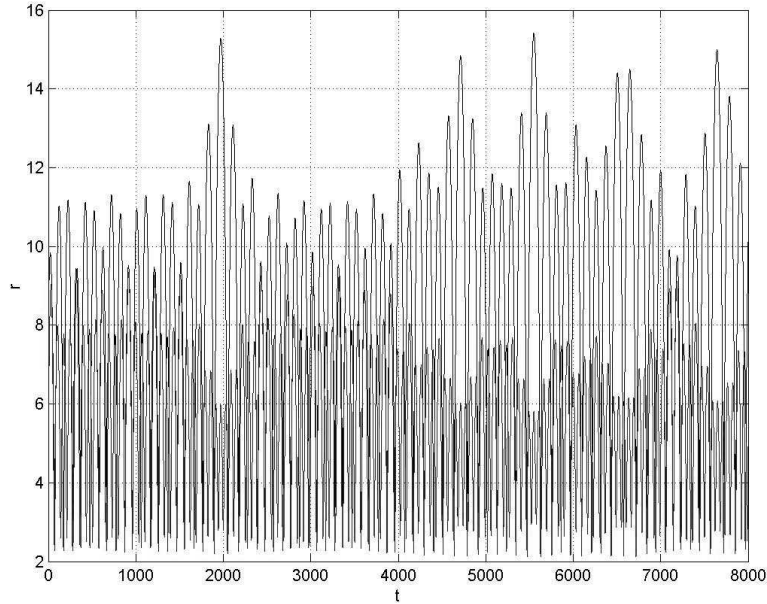,width=12cm}}
\caption{
 Chaotic shell oscillations with mass ratio
$m/M=0.15$.}
\label{fig3}
\end{figure}

Calculations by Barkov et al. (2001) have shown, that very small
variations in the initial parameters drastically
change the picture of the oscillations, which is a characteristic for
a chaotic behavior.
In the case of massive shells the exchange of energy between
shells occurs more intensively. As a result we have the obviously
expressed chaotic behavior of shells presented
in fig.\ref{fig3} from Barkov et al. (2001) for the case
$m/M=0.15$.

\begin{figure}
\centerline{\epsfig{figure=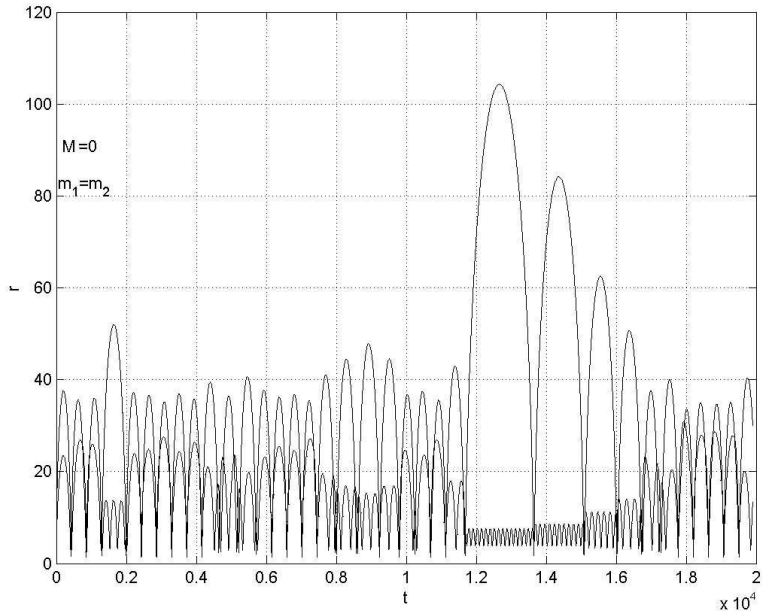,width=12cm}}
\caption{
 The chaotic motion $r_{1,2}(t)$ of two
self-gravitating shells in their own gravitational field without a
central mass.}
\label{fig4}
\end{figure}

The example of a chaotic behavior of two intersecting
self-gra\-vi\-ta\-ting shells, moving in their own gravitational field,
without a central mass, is shown in fig. \ref{fig4}.
Some other example of the chaotic motion of intersecting
self-gravitating shells are given by Barkov et al. (2001).

Ballistic mechanism of energy exchange between gravitating
particles may be important in the
formation of structures in a
 cold dark matter, where gravitational instabilities
are developing in masses, much less than galactic ones, and
gravitationally bound massive objects, consisting mainly from the
dark matter may be formed.

\section{Stars around a supermassive black hole in active galactic
nuclei}

It is now widely believed that active galactic nuclei and quasars
are radiating due to accretion of matter into a supermassive black
hole, according to Lynden-Bell (1969) model. A supermassive black
hole is surrounded by a dense stellar cluster. Its member stars
may supply matter for the accretion, when they make a close
approach to the central body. The fate of the star depends on the
mass of the black hole. When $M_{bh}< 3\times 10^7 M_{\odot}$, the
solar-type stars are destroyed by tidal forces at radius
$r_t=(M_{bh}/M_{\odot})^{1/3} R_{\odot}$, and matter is forming an
accretion disk. For
black holes with mass greater than $3{\times}10^{7}M_{\odot}$
tidal forces are
not enough for disruption of such stars, and they are falling
directly into the black hole from the distance of $2-3$ the
Schwarzschild gravitational radius $r_g=2GM_{bh}/c^2$. This is
because the radius of the tidal disruption $r_t$ depends on the
black hole mass as $r_t \sim M_{bh}^{1/3}$, and the radius of the
gravitational capture $r_{grav} \sim M_{bh}$, so at large $M_{bh}$
we have $r_t<r_{grav}$, and gravitational capture occurs before
the tidal disruption happens
 (see e.g. Bisnovatyi-Kogan et al., 1980).
 The tidal disruption of stars may be an important supply of matter
 in low-luminosity AGN (Rees, 1994).

The region in the phase space from which stars may be disrupted or
absorbed by the black hole occupies a small region called loss
cone. In the non-collisional stellar cluster this cone is almost
empty, and it is filled only due to rare collisions which provide
a diffusion of stars into a loss cone. This process may be very
slow, and sometimes cannot give enough matter to explain the most
luminous objects. The situation may be more optimistic when stars
around the black hole move along chaotic orbits due to the particular
shape of a gravitational potential. Filling of
the loss cone, according to Norman and Silk (1983),
occurs not only by slow diffusion, but also by stars entering
the loss cone during the motion along chaotic orbits.
The latter may even be more effective.

\section{Stellar oscillations}

\begin{figure}
\centerline{\epsfig{figure=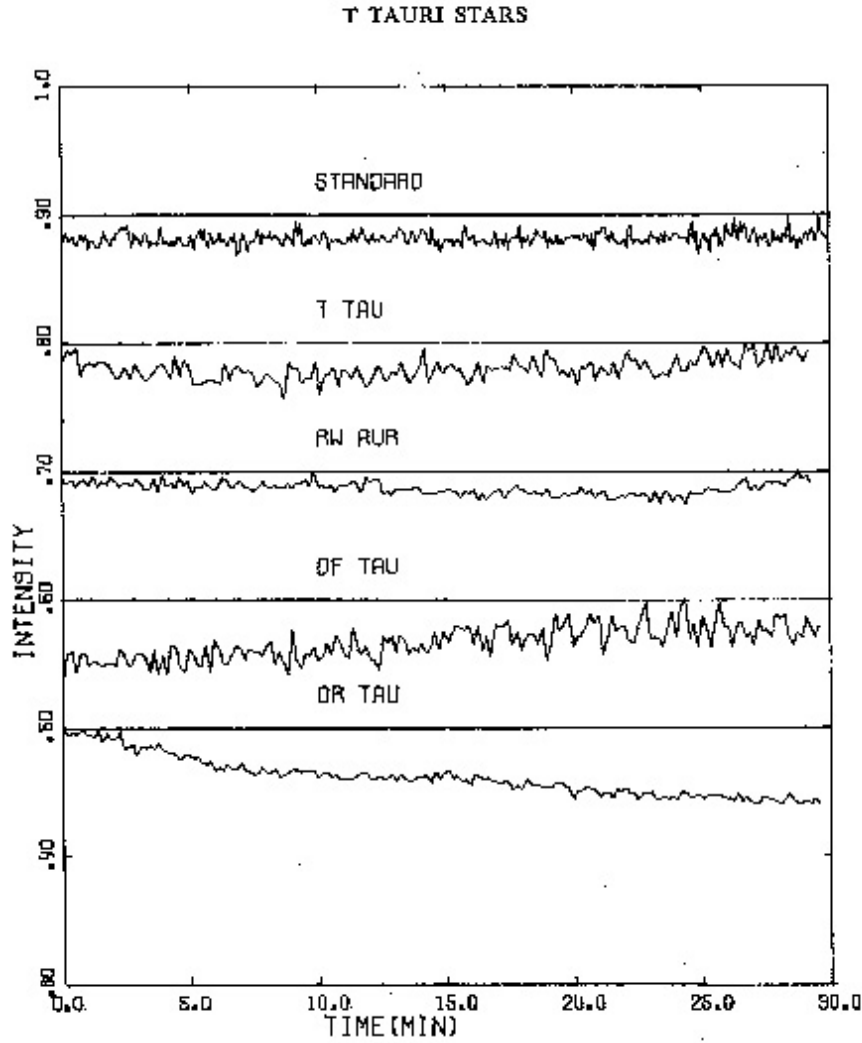,width=12cm}}
%\centerline{\epsfig{figure=fkuan.ps,width=12cm}}
\caption
{Time scans of the ultraviolet intensities  of a standard star
and several T Tauri stars obtained on 1976 January 26 (UT). The
integration time for the standard is 5 s and for the rest 10 s.
The intensities have been corrected for atmospheric extinction and
are in arbitrary unit. } \label{fig5}
\end{figure}

\begin{figure}
\centerline{\epsfig{figure=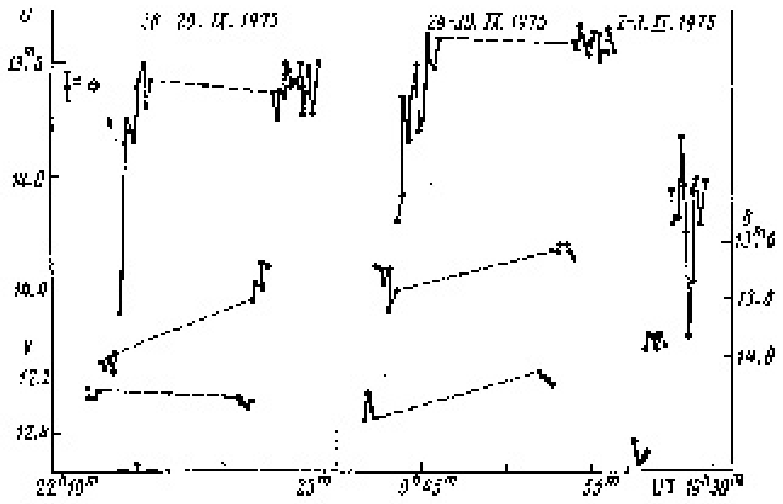,width=7cm}\,\,
            \epsfig{figure=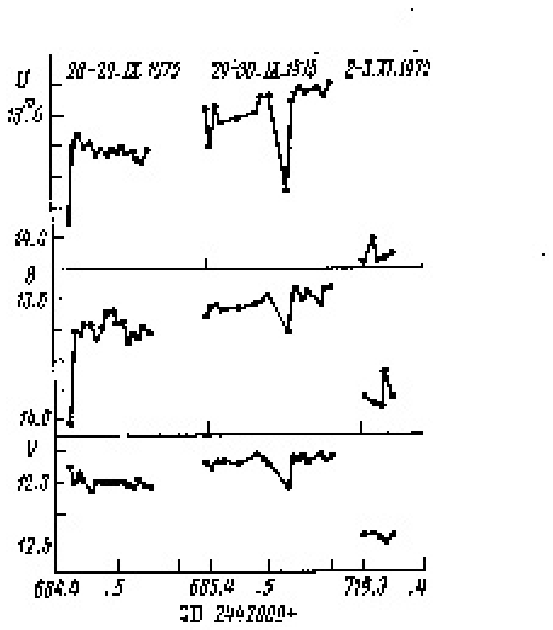,width=5cm}\,\,}
%\centerline{\epsfig{figure=fzl1.ps,width=8cm}\,\,
%            \epsfig{figure=fzl1.ps,width=6cm}\,\,           }
\caption{Photo-electric light curves of DF Tau in the U, B, V
system. Each
         dot corresponds to a 10-sec integration time. The gaps in the
         light curves represent observations of a comparison star and
         the sky background ({\it left}).
Light curves of DF Tau. Each point corresponds to an integration time
of 80-100 sec in U and 40-50 sec in B and V
 ({\it right}).}
\label{fig6}
\end{figure}

Stars show all kinds of variability: from periodic to totally irregular. The
variability arises from the influence of the thermal processes and
displays in
 their dynamic behavior. Irregular variability is
observed in stars with large convective envelopes. The most
intensive convection is developed in the envelopes of cool stars,
where matter opacity is high, and radiative gradient exceeds the
adiabatic one. This is related to stars with masses 1-2
$M_{\odot}$ in the stage of gravitational contraction (T Tauri
stars), and to low-mass stars with $M \le \sim 0.3 M_{\odot}$,
which remain almost fully convective during all their life. Both
type of stars show irregular variability in the form of long and
short flares. Short time variability of of several T Tauri type
star is represented in fig. \ref{fig5} from the paper of Kuan
(1976); the light curve of T Tauri type star DF Tauri in different
time scales is shown in fig.\ref{fig6} from the paper of
Zaitseva and Lyutyi (1976).

 Detailed description of
chaotic flares in low mass UV Ceti stars
is given in the book of Gershberg (1978).
Light curves  in figs. \ref{fig5}, \ref{fig6}, show an evident
chaotic behavior.

\begin{figure}
\centerline{\epsfig{figure=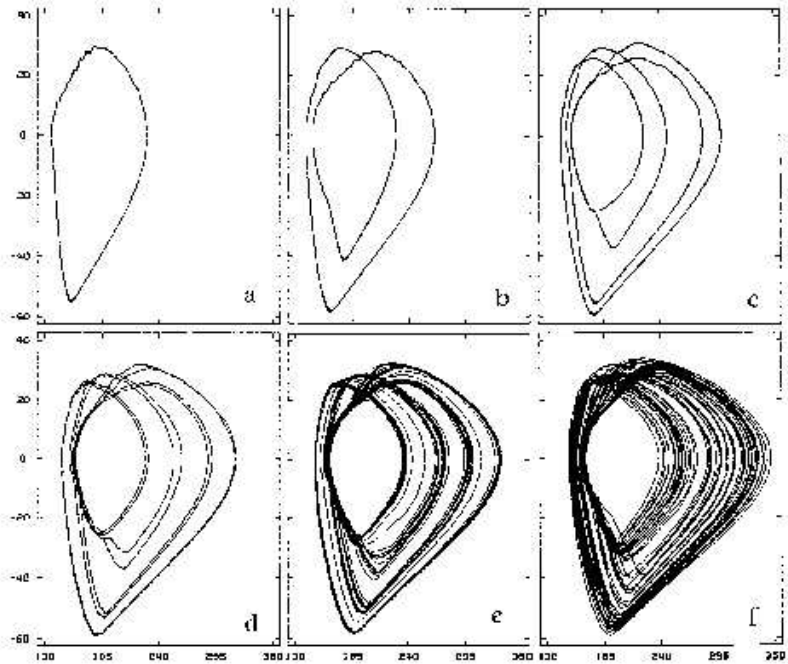,width=12cm}}
%\centerline{\epsfig{figure=fbuchko.ps,width=12cm}}
 \caption{Plots
of $v_*$ (ordinate) - $R_*$ (abscissa) for models $a$ trough $f$
(the asterisk denotes the 55th zone of the models); $R_*$ in
$10^{10}$ cm, $v_*$ in km s$^{-1}$. The logarithms of effective
temperature $T_{ef}$, in $K$, $\log T_{ef}$ (and corresponding
periods ) are the following: $a$ - 3.71 (11.587); $b$ - 3.69
(26.196); $c$ - 3.67 (59.040); $d$ - 3.66 (125.784); $e$ - 3.65
($\infty$); $f$ - 3.64 ($\infty$). } \label{fig7}
\end{figure}

Another type of chaos in stellar light curve appears in pulsating
stars in which the regular pulsations are transforming into the
chaotic ones in the course of stellar evolution. The transition
from regular pulsations to the chaotic regime occurs
with change of a characteristic during the evolution, in agreement
with the mechanism of period
doubling (Feigenbaum,
1983). The behavior becomes purely chaotic when the parameter
$\lambda$ is approaching the limiting value $\lambda_{lim}$. The
universal law describing the approach to chaos was discovered by
Feigenbaum (1983). He had found that values of the parameters
$\lambda_n$ at which a corresponding period doubling happens,
follow the limiting relation

\begin{equation}
\label{eq7}
\lim \frac{\lambda_{n+1}-\lambda_{n}}{\lambda_{n+2}-\lambda_{n+1}}=
4.6692016....
\end{equation}
This law was checked in different physical, mechanical and pure
mathematical examples. For pulsating stars this law was confirmed
quantitatively for a few
 W Vir and RV Tau type stars.
The example of such W Vir type star with luminosity $L=400
L_{\odot}$, mass $M=0.6 M_{\odot}$, and effective temperature
$T_{ef}$ changing from  $\log T_{ef}=3.71$ until 3.64 was analyzed
by Buchler and Kovacs (1987). They obtained that transition to
chaos happens at $\log T_{ef}=3.65$, when the motion became
aperiodic and chaotic. Fig. \ref{fig7} from the paper of Buchler
and Kovacs (1987) illustrates the transition from the periodic to
chaotic behavior via the period doubling mechanism.

\section{High-energy sources}

Irregular variabilities are observed in most X-ray and gamma-ray
so\-urces. Strong X-ray so\-urces consist of a neutron star or a
black hole in a binary system, and the main energy supply in these
objects comes from accretion into the compact star from the
companion. In the case of a black hole the main source of
radiation is an accretion disk, which is formed due to high
angular momentum of the falling matter. In the case of a neutron
star, in addition to the accretion disks,
important processes occur on the surface of the neutron star,
and in its magnetosphere, when the star is strongly magnetized.
Both the accretion disk and the neutron star surface, as well as the
magnetosphere suffer from different kind of instabilities, leading
to irregular radiation flux. The instabilities in the accretion
disk of visco-thermal origin determine short-time fluctuations on
scales from seconds to milliseconds, observed in most X-ray
sources, containing black holes (Cherepashchuk, 1996). Accretion
disk instabilities, determining transition between quasi-laminar
and highly turbulent states are probably responsible for
appearance of soft X-ray transient (X-ray novae) (see review of
Cherepashchuk (2001)). Similar instability may also explain
cataclismic variables (see Spruit and Taam, 2001), where flashes
occur with intervals of few months. These sources are also binary
systems, containing white dwarf and low-mass star supplying matter
for an accretion disk formation around the white dwarf. The energy
of these flashes is several orders of magnitude less than in X-ray
novae.

Interaction of matter with stellar surface and stellar
magnetosphere determine chaotic features in radiation of the
sources, containing neutron stars. Voges et al. (1987) have found
irregular variability in the form of "deterministic chaos" in the
variations of the pulse shape of the famous X-ray pulsar Her X-1.
It is probably connected with instabilities in the accretion flow
inside the magnetosphere.

\begin{figure}
\centerline{\rotatebox{-90}{\epsfig{figure=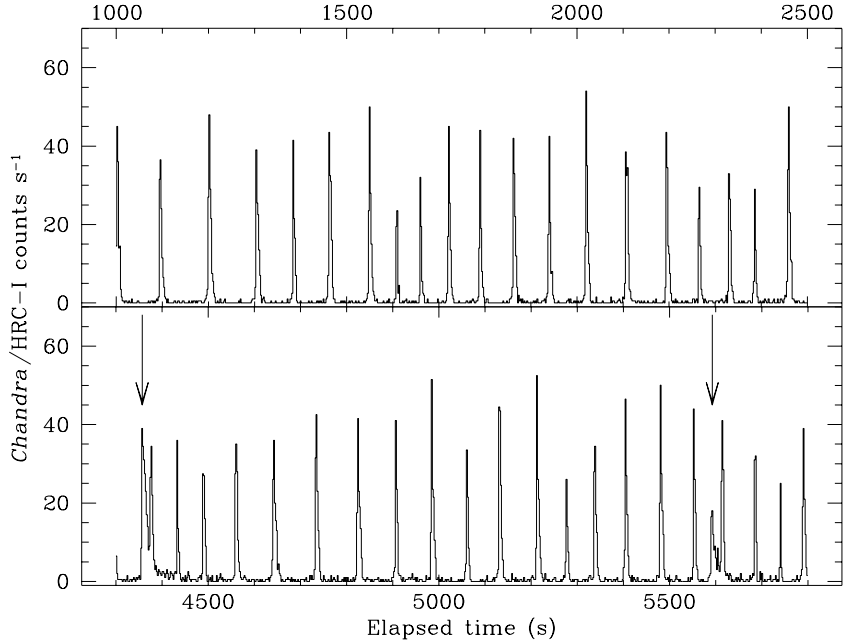,width=9cm}}}
\caption{ Illustrative 1500s-long sections of the {\it
Chandra}/HRC-I light curve of the Rapid Burster, with 1s binning
applied.  Throughout the $\sim$13 ks observation the source shows
regular type-II bursts (top panel); in addition two type-I burst
event have been detected (bottom panel, indicated by arrows). }
\label{fig8}
\end{figure}

Matter falling into the neutron star becomes degenerate soon after
joining its envelope, when pressure almost does not depend on the
temperature. In this conditions the thermal instability of
thermonuclear burning develops, leading to appearance of X-ray
bursters, observed in low-mass X-ray binary systems. This bursts
occur non-periodically and show chaotic properties. Example of
chaotic flares in the unique famous burster source, called "rapid
burster" is represented in fig \ref{fig8} from the paper of Homer et
al. (2001) using observations of Chandra X-ray satellite.

\section{Fractal structure of detonation front in SN I}

\begin{figure}
\centerline{\epsfig{figure=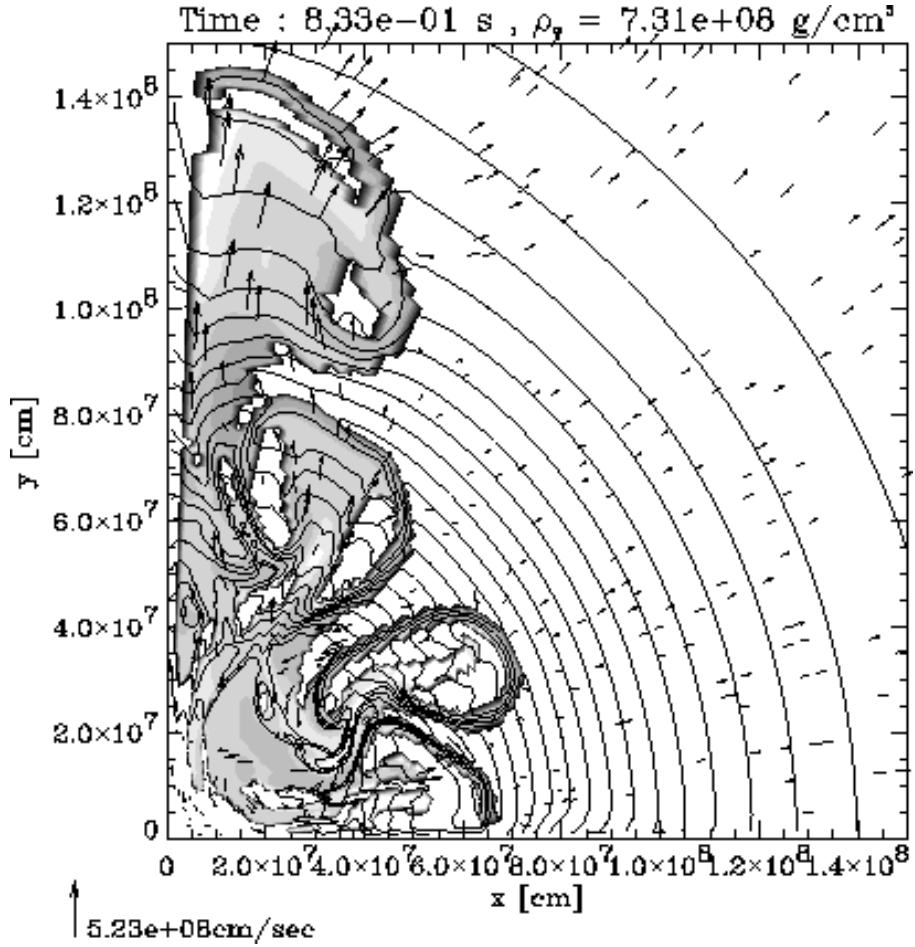,width=12cm}}
\caption{
 Core region in the SN I - model at maximally developed RT instability
Density contours are separated by $\delta \rho = 3 \cdot 10^7$ g
cm$^{-3}$, the maximum energy generation rate (shaded regions) is
$\dot S_{\rm max} = 1.3 \cdot 10^{19}$ ergs g$^{-1}$ s$^{-1}$. }
\label{fig9}
\end{figure}

Supernovae explosions are the most spectacular astronomical
events, by which massive stars end their life. Type I SN appears
due to thermonuclear explosion of a degenerate $C-O$ stellar core
with a mass equal to the Chandrasekhar mass limit $\sim 1.4\,
M_{\odot}$. The thermal instability starts in the central part of
the core, and the flame propagates outside to the surface of the
core. Physical and astronomical analysis of the problem led to
conclusion, that the flame propagation should be highly unstable,
and the surface of the front of the flame should be  highly
non-smooth, with a fractal properties. Development of a fractal
structure of the flame front strongly
enhances the surface of the
burning, increasing the rate of the energy production. That may
change the regime of the flame propagation from deflagration, when
the flame front moves with a sub-sonic speed, to detonation, in
which this front moves with the speed equal or larger than the
local sound speed. The necessity of this transition follows also
from the interpretation of observational data on the isotope
distribution of different elements, produced by ejection of matter
during SN I explosions. Theoretical analysis have shown existence
of two types of instabilities in the flame front: Landau-Darrieus
(Blinnikov and Sasorov, 1996), and Rayleigh-Taylor (RT)
instability (Niemeyer et al., 1997). 2-D numerical calculations
made by Niemeyer et al. have shown a development of RT
instability, which appeared in many scales, approaching the
fractal structure (see fig. \ref{fig9}).

The simulations involved an Euler\-ian PPM-based code to solve the
2-D hydro-dy\-na\-mical equations.
 All calculations were
based on a 2-D stationary grid with $256 \times 64$ zones in
spherical ($r, \vartheta$) coordinates. Assuming rotational and
equatorial symmetry, the boundary conditions were chosen to be
reflecting everywhere except at the outer radial edge, where
outflow was allowed. The initial model represented  a white dwarf
with a mass equal to the mass of the Chandrasrkhar limit, a
central density $\rho_{\rm c} = 2.8 \times 10^9$ g cm$^{-3}$ and a
central temperature $T_{\rm c} = 7 \times 10^8$ K. At $t \approx
0.8$ s, displayed in fig.~\ref{fig9}, the RT-instability of the
burning front has developed to its maximum extent. The turbulent
flame speed is now $u(\Delta) \approx 2 \times 10^7$ cm/s at the
points of maximum turbulent sub-grid energy. Four major bubbles
can be identified, separated by thin streams of unburned
C+O-material.

\section{Stellar dynamos}

\begin{figure}
\centerline{\epsfig{figure=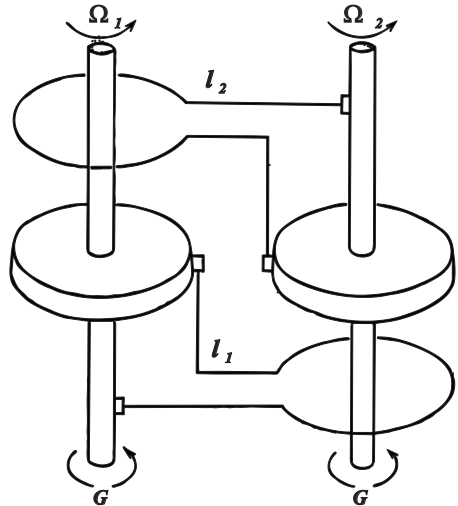,width=6cm}}
\caption{
A schematic picture of Rikitake dynamo.}
\label{fig10}
\end{figure}

Dynamo models are used for explanation of the origin of stellar
and planet magnetic fields. The irregular changes in the Earth
magnetic field polarity with intervals $\Delta t\,= \,
(0.1\,-\,20)\times 10^6$ years have been explained by the chaotic
behavior of the dynamo process. The famous example of the chaotic
dynamo is the toy two-disk Rikitake model (Cook and Roberts,
1970).
 This model consists of two magnetically connected disks (see
fig. \ref{fig10}) with the following given parameters:

$G$ - momentum of force,
$\Omega$ - angular velocity,
$I$ - electrical currant,
$R$ - resistivity,
$L$ - self-inductance,
$M$ - mutual inductance,
$C$ - momentum of inertia.

The equation describing the behavior of this system (mechanical and
magnetic) are written as

\begin{equation}
\label{eq8}
L_1 \frac{dI_1}{dt}\,=\,-RI_1+M \Omega_1 I_2, \qquad
C_1 \frac{d\Omega_1}{dt}\,= \, G_1-M I_1 I_2
\end{equation}
for the first disk, and corresponding equations for the second disk.
Taking identical disks under the same external action $G$, we find two
values of time, characterizing the system:

\begin{equation}
\label{eq10}
\tau_m=\frac{CR}{GM}, \quad {\rm and }\quad \tau_l=\frac{L}{R}.
\end{equation}
Here $\tau_m$ estimates the acceleration time, and $\tau_l$ characterizes
the magnetic field damping. The ratio of these times $\mu^2=\tau_M/\tau_l$
is the non-dimensional parameter dividing stochastic behavior ($\mu\ge 1$)
from the regular oscillations.

There are indications that the solar 22 year cycle is also showing
stochastic behavior on long-term period, and could be considered as a
coherent strange attractor (Ruzmaikin et al., 1992).

\section {Chaos in Solar system}

Appearance of most comets in the region of visibility is
unpredictable. Theoretical analysis of Sagdeev and Zaslavsky
(1987) had shown, that gravitational influence of Jupiter and
Saturn creates a layer around the Sun in which orbits of comets
become stochastic. Diffusion of the orbit parameters of comets in
the stochastic layer make them to enter randomly the region when
they become visible.

\acknowledgements The work was partially supported by RFFI grant
99-02-18180, INTAS-ESO grant No.120 and INTAS grant 00-491.

\theendnotes

\end{article}
\end{document}